\documentclass[twocolumn,showpacs,preprintnumbers,amssymb,nofootinbib]{revtex4}

\usepackage{graphicx}
\usepackage{bm} 

\begin{document}


\title{Pair creation of anti-de Sitter black holes on a cosmic string background}

\author{\'Oscar J. C. Dias}
\email{oscar@fisica.ist.utl.pt} \affiliation{ Centro
Multidisciplinar de Astrof\'{\i}sica - CENTRA, \\Departamento de
F\'{\i}sica, Instituto Superior T\'ecnico, \\Av. Rovisco Pais 1,
1049-001 Lisbon }

\date{\today}
\begin{abstract}
 We analyze the quantum process in which a cosmic
string breaks in an anti-de Sitter (AdS) background, and a pair of
charged or neutral black holes is produced at the ends of the
strings. The energy to materialize and accelerate the pair comes
from the strings tension. In an AdS background this is the only
study done in the process of production of a pair of correlated
black holes with spherical topology. The acceleration $A$ of the
produced black holes is necessarily greater than
$\sqrt{|\Lambda|/3}$, where $\Lambda<0$ is the cosmological
constant. Only in this case the virtual pair of black holes can
overcome the attractive background AdS potential well and become
real. The instantons that describe this process are constructed
through the analytical continuation of the AdS C-metric. Then, we
explicitly compute the pair creation rate of the process, and we
verify that (as occurs with pair creation in other backgrounds)
the pair production of nonextreme black holes is enhanced relative
to the pair creation of extreme black holes by a factor of
$e^{{\cal A}_{\rm bh}/4}$, where ${\cal A}_{\rm bh}$ is the black
hole horizon area. We also conclude that the general behavior of
the pair creation rate with the mass and acceleration of the black
holes is similar in the AdS, flat and de Sitter cases, and our AdS
results reduce to the ones of the flat case when $\Lambda
\rightarrow 0$.
\end{abstract}


\maketitle

\section{\label{sec:Int}Introduction}
The quantum Schwinger-like process of black hole pair creation in
an external field is by now a well studied subject. It supplied a
way to produce Planck scale black holes, and it gave useful clues
to the discussion of issues like the statistical properties of
black holes, the black hole information paradox, and the analysis
of topology changing processes.

In order to turn the pair of virtual black holes real one needs a
background field that provides  the energy needed to materialize
the pair, and that furnishes the force necessary to accelerate
away the black holes once they are created. This background field
can be: (i) an external electromagnetic field with its Lorentz
force (see \cite{Gibbons-book}-\cite{Emparan}), (ii) the positive
cosmological constant $\Lambda$, or inflation (see
\cite{MelMos}-\cite{BoussoDil}), (iii) a cosmic string with its
tension (see \cite{HawkRoss-string}-\cite{PreskVil}), (iv) a
domain wall with its gravitational repulsive energy (see
\cite{CaldChamGibb}-\cite{Rogatko}). One can also have a
combination of the above fields, for example, a process involving
cosmic string breaking in a background magnetic field
\cite{Empar-string}, or a scenario in which a cosmic string breaks
in a positive cosmological background \cite{OscLem-PCdS}.

The evaluation of the black hole pair creation rate has been done
at the semiclassical level using the instanton method.  An
instanton is an Euclidean solution that interpolates between the
initial and final states of a classically forbidden transition,
and is a saddle point for the Euclidean path integral that
describes the pair creation rate.
 The instantons that mediate the above processes can be
 obtained by analytically continuing (i) the Ernst solution \cite{Ernst}
(which is available only for $\Lambda=0$), (ii) the de Sitter
black hole solutions, (iii) the C-metric with a generalized
cosmological constant $\Lambda$ \cite{KW,PlebDem}, or (iv) the
domain wall solution \cite{VilenkinStringIpserSikivie}. All these
exact solutions describe appropriately the evolution of the black
hole pair that has been created, since they represent a pair of
black holes accelerated by an electromagnetic field, by a
cosmological constant, by a string, or by a domain wall,
respectively.

An important process that accompanies the production of the black
hole pair  is the emission of electromagnetic and gravitational
radiation. In a flat background, an estimate for the amount of
gravitational radiation released during the pair creation period
has been given in \cite{VitOscLem}. After the pair creation, the
black hole pair accelerates away and, consequently, the black
holes continue to release gravitational and electromagnetic
energy. In a $\Lambda=0$ background, the gravitational radiation
emitted by uniformly accelerated black holes has been computed in
\cite{BicPravPrav}, while in a dS background this analysis has
been carried in \cite{BicKrtKrtPod}, and in an AdS background the
programme has been completed in \cite{PodOrtKrtAdS}. We ask the
reader to see, e.g., the introduction of \cite{OscLem-PCdS} for a
recent and detailed description of the works done on black hole
pair creation processes.

In this paper we want to analyze the process in which a cosmic
string breaks and a pair of black holes is produced at the ends of
the string, in an anti-de Sitter (AdS) background ($\Lambda<0$).
Therefore, the energy to materialize and accelerate the pair comes
from the strings tension. In an AdS background this is the only
study done in the process of production of a pair of correlated
black holes with spherical topology. The analysis of this process
in a flat background ($\Lambda=0$) has been done in
\cite{HawkRoss-string}, while in a de Sitter background
($\Lambda>0$) it has been carried in
\cite{OscLem-PCdS,OscLem_dS-C}). The instantons for this process
can be constructed by analytically continuing the AdS C-metric
found in \cite{PlebDem} and analyzed in detail in
\cite{EHMP,OscLem_AdS-C}. Contrary to the $\Lambda=0$ \cite{KW}
and $\Lambda>0$ \cite{OscLem_dS-C} cases, the AdS describes a pair
of accelerated black holes only when the acceleration supplied by
the strings is greater than $\sqrt{|\Lambda|/3}$
\cite{OscLem_AdS-C}. Hence we expect that pair creation of black
holes in an AdS background is possible only when
$A>\sqrt{|\Lambda|/3}$. We will confirm this expectation. The
quantum production of uncorrelated AdS black holes has been
studied in \cite{WuAdS}, and the pair creation process of
topological AdS black holes (with hyperbolic topology) has been
analyzed in \cite{MannAdS} in a domain wall background.

The plan of this paper is as follows. In Sec. \ref{sec:AdS C-inst}
we construct, from the AdS C-metric, the instantons that describe
the pair creation process. In Sec. \ref{sec:Calc-I AdS}, we
explicitly evaluate the pair creation rate of black holes in an
AdS background when a string breaks. In Sec. \ref{sec:Conc AdS}
concluding remarks are presented. In the Appendix \ref{sec:A1}, we
discuss, for the benefit of comparison with the $\Lambda=0$ case,
the pair creation of black holes in a $\Lambda=0$ background. We
explicitly compute the numerical value of the pair creation rate,
which has not been done yet. In the Appendix \ref{sec:A2} a
heuristic derivation of the pair creation rates is given.
Throughout this paper we use units in which $G=c=\hbar=1$.

\section{\label{sec:AdS C-inst} The A\lowercase{d}S C-metric instantons}

The AdS C-metric has been found in \cite{PlebDem}. The physical
properties and interpretation of this solution have been analyzed
in \cite{OscLem_AdS-C,EHMP}. When $A>\sqrt{|\Lambda|/3}$, and only
in this case \cite{OscLem_AdS-C}, the AdS C-metric describes a
pair of uniformly accelerated black holes in an AdS background,
with the acceleration being provided by two strings, from each one
of the black holes towards infinity, that pulls them away. Since
we are interested in black hole pair creation, hereafter we will
deal only with the $A>\sqrt{|\Lambda|/3}$ case. For a detailed
discussion on the properties of the AdS C-metric we ask the reader
to see \cite{OscLem_AdS-C}. Here we will only mention those which
are really essential.

\begin{figure} [t]
\includegraphics*[height=2.2in]{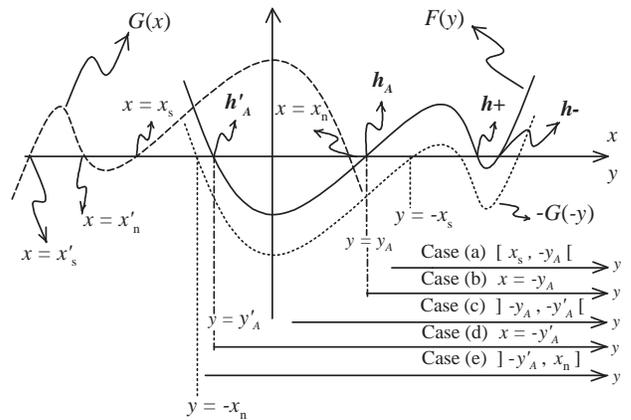}
\caption{\label{g3_pc_AdS}
 Shape of ${\cal G}(x)$ and ${\cal F}(y)$ for a general
nonextreme charged massive AdS C-metric. The allowed range of $x$
is between $x_\mathrm{s}$ and $x_\mathrm{n}$ where ${\cal G}(x)$
is positive and compact. In the Lorentzian sector, the permitted
range of $y$ is $-x\leq y < +\infty$. The presence of an
accelerated horizon is indicated by $y_A$ and $y'_A$ and the inner
and outer charged horizons by $y_-$ and $y_+$. In the extreme case
$y_-$ and $y_+$ coincide, and ${\cal G}(x)$ has only two real
roots represented by $x_\mathrm{s}$ and $x_\mathrm{n}$. When $m=0$
and $q=0$, both ${\cal G}(x)$ and ${\cal F}(y)$ are represented by
a parabola with the only zeros of ${\cal G}(x)$ being
$x_\mathrm{s}$ and $x_\mathrm{n}$, and the only zeros of
 ${\cal F}(x)$ being $y'_A$ and $y_A$.
 }
\end{figure}

The gravitational field of the Lorentzian AdS C-metric is given by
(see, e.g., \cite{OscLem_AdS-C})
\begin{equation}
 d s^2 = [A(x+y)]^{-2} (-{\cal F}dt^2+
 {\cal F}^{-1}dy^2+{\cal G}^{-1}dx^2+
 {\cal G}d\phi^2)\:,
 \label{C-metric}
 \end{equation}
with
 \begin{eqnarray}
 & &{\cal F}(y) = -\frac{3A^2-|\Lambda|}{3A^2}
                     +y^2-2mAy^3+q^2A^2y^4, \nonumber \\
 & &{\cal G}(x) = 1-x^2-2mAx^3-q^2 A^2 x^4\:,
 \label{FG PCAdS}
 \end{eqnarray}
where $\Lambda<0$ is the cosmological constant, $A>0$ is the
acceleration of the black holes, and $m$ and $q$ are the ADM mass
and electromagnetic charge of the non-accelerated black hole,
respectively. The Maxwell field in the magnetic case is given by
\begin{eqnarray}
 F_{\rm mag}=-q\, dx\wedge d\phi \:,
\label{F-mag}
\end{eqnarray}
while in the electric case it is given by
 \begin{eqnarray}
F_{\rm el}=-q\, dt \wedge dy \:.
 \label{F-el}
\end{eqnarray}
The general shape of ${\cal G}(x)$ and ${\cal
 F}(y)$ is represented in Fig. \ref{g3_pc_AdS}.

The solution has a curvature singularity at $y=+\infty$ where the
matter source is.  The point $y=-x$ corresponds to a point that is
infinitely far away from the curvature singularity, thus as $y$
increases we approach the curvature singularity and $y+x$ is the
inverse of a radial coordinate. At most, ${\cal F}(y)$ can have
four real zeros which we label in ascending order by $y_{\rm
neg}<0<y_A\leq y_+ \leq y_-$. The roots $y_-$ and $y_+$ are
respectively the inner and outer charged black hole horizons, and
$y_A$ is an acceleration horizon. The negative root $y_{\rm neg}$
satisfies $y_{\rm neg}<-x$ and thus has no physical significance.
The angular coordinate $x$ belongs to the range
$[x_\mathrm{s},x_\mathrm{n}]$ for which ${\cal G}(x)\geq 0$ (when
we set $A=0$ we have $x_\mathrm{s}=-1$ and $x_\mathrm{n}=+1]$). In
order to avoid a conical singularity in the north pole, the period
of $\phi$ must be given by
\begin{equation}
\Delta \phi=\frac{4 \pi}{|{\cal G}'(x_\mathrm{n})|}\:,
 \label{Period phi}
 \end{equation}
and this leaves a conical singularity in the south pole with
deficit angle
\begin{eqnarray}
\delta =2\pi \left ( 1-\frac{{\cal G}'(x_\mathrm{s})}{|{\cal
G}'(x_\mathrm{n})|} \right )\,.
 \label{conic-sing}
 \end{eqnarray}
that signals the presence of strings from each one of the black
holes towards infinity with mass density $\mu =\delta/(8\pi)$, and
with pressure $p=-\mu<0$. When we set the acceleration parameter
$A$ equal to zero, the AdS C-metric reduces to the usual
AdS$-$Reissner-Nordstr\"{o}m or AdS-Schwarzschild solutions
without conical singularities.

Later on it will be crucial to note that the number and nature of
the horizons crossed by an observer that travels into the black
hole singularity depends on the angular direction $x$ that he is
following (see Fig. \ref{g3_pc_AdS}). This peculiar feature is due
to the lower restriction on the value of $y$ ($-x\leq y$), and a
detailed discussion and explanation of it can be found in
\cite{OscLem_AdS-C} (such an angular dependence occurs also in the
flat C-metric \cite{KW}, although not so sharply). For example,
when the observer is travelling towards the black hole singularity
following an angular direction in the vicinity of the south pole
($x_\mathrm{s}\leq x <-y_A$), he will cross only the outer $y_+$
and inner $y_-$ black hole horizons (see Fig. \ref{g3_pc_AdS}),
while when he does this trip following an angular direction in the
vicinity of the equator ($ -y_A < x <-y'_A$), he crosses the
acceleration horizon $y_A$ before passing through the black hole
horizons $y_+$ and $y_-$.

In order to evaluate the black hole pair creation rate we need to
find the instantons of the theory, i.e., we must look into the
Euclidean section of the AdS C-metric and choose those Euclidean
solutions which are regular in a way that will be explained soon.
The Euclidean section of the AdS C-metric is obtained from the
Lorentzian AdS C-metric by introducing an imaginary time
coordinate $\tau=-it$ in (\ref{C-metric})-(\ref{F-el}). To have a
positive definite Euclidean metric we must require that $y$
belongs to $y_A \leq y \leq y_+$. In general, when $y_+ \neq y_-$,
one then has conical singularities at the horizons $y=y_A$ and
$y=y_+$. In order to obtain a regular solution we have to
eliminate the conical singularities at both horizons, ensuring in
this way that the system is in thermal equilibrium. This is
achieved by imposing that the period of $\tau$ is the same for the
two horizons, and is equivalent to requiring that the Hawking
temperature of the two horizons be equal. To eliminate the conical
singularity at $y=y_A$ the period of $\tau$ must be $\beta=2 \pi/
k_A$ (where $k_A$ is the surface gravity of the acceleration
horizon),
\begin{equation}
\beta=\frac{4 \pi}{|{\cal F}'(y_A)|}\:.
 \label{Period tau-yA}
 \end{equation}
 This choice for the period of $\tau$ also eliminates
simultaneously the conical singularity at the outer black hole
horizon, $y_+$, if and only if the parameters of the solution are
such that  the surface gravities of the  black hole and
acceleration horizons are equal ($k_+=k_A$), i.e.
\begin{equation}
 {\cal F}'(y_+)=-{\cal F}'(y_A)\:.
 \label{k+=kA}
 \end{equation}
This condition is satisfied by a regular Euclidean solution with
$y_A \neq y_+$ that will be referred to as nonextreme AdS
instanton.

We now turn our attention to the case $y_+ = y_-$ (and $y_A\neq
y_+$). When this happens the allowed range of $y$ in the Euclidean
sector is simply $y_A \leq y < y_+$. This occurs because when $y_+
= y_-$ the proper distance along spatial directions between $y_A$
and $y_+$ goes to infinity. The point $y_+$ disappears from the
$\tau, y$ section which is no longer compact but becomes
topologically $S^1 \times {\mathbb{R}}$. Thus, in this case we
have a conical singularity only at $y_A$, and so we obtain a
regular Euclidean solution by simply requiring that the period of
$\tau$ be equal to (\ref{Period tau-yA}). We will label this
solution by extreme AdS instanton.

In a de Sitter (dS) background there is another $y_+ \neq y_-$
instanton  which satisfies $y_A=y_+$. It is called Nariai
instanton. Moreover, in the dS background, there is also a special
solution that satisfies
 $y_A=y_+=y_-$. It is called ultracold instanton. In the AdS
 C-metric case, the counterparts of these dS instantons are of no
 interest for the pair creation process because they are out of
 the allowed range of the angular direction $x$.

In Secs. \ref{sec:Lukewarm-inst AdS} and
 \ref{sec:Cold-inst AdS} we will describe in detail
the nonextreme AdS instanton with $m=q$ and the extreme AdS
instanton with $y_+=y_-$, respectively. These instantons are the
natural AdS C-metric counterparts of the lukewarm dS C and cold dS
C instantons constructed in \cite{OscLem_dS-C}. Thus, these
instantons could also be labelled as lukewarm AdS C and cold AdS C
instantons. Moreover, in Sec. \ref{sec:submaximal-inst AdS}, we
will also describe saddle point solutions that have conical
singularities. These solutions represent nonextreme instantons
with $m\neq q$. In Sec. \ref{sec:background string AdS} we will
study the background reference spacetime that describes the
initial system before the pair creation. These results will allow
us to calculate the pair creation rate of accelerated AdS black
holes in Sec. \ref{sec:Calc-I AdS} .

\subsection{\label{sec:Lukewarm-inst AdS}The nonextreme AdS
instanton with $\bm{m=q}$}

As we said above, for the nonextreme AdS instanton the
gravitational field is given by (\ref{C-metric}) with the
requirement that ${\cal F}(y)$ satisfies ${\cal F}(y_+)=0={\cal
F}(y_A)$ and ${\cal F}'(y_+)=-{\cal F}'(y_A)$. In this case we can
then write
\begin{eqnarray}
{\cal F}(y)&=&-\left ( \frac{y_A \: y_+}{y_A+y_+} \right )^2 \left
( 1-\frac{y}{y_A} \right ) \left ( 1-\frac{y}{y_+} \right )
\nonumber\\
 & & \times
\left ( 1+\frac{y_A+y_+}{y_A \:y_+}\,y-\frac{y^2}{y_A \:y_+}
\right
 )\:,
 \label{F-luk PCAdS}
 \end{eqnarray}
with
\begin{eqnarray}
& & y_A =  \frac{1-\alpha}{2mA}\,, \:\:\:\:\:\:\:\: y_+ =
\frac{1+\alpha}{2mA}\,, \nonumber\\
 & & {\rm and} \:\:\:
 \alpha = \sqrt{1-\frac{4m}{\sqrt{3}}\sqrt{3A^2-|\Lambda|}} \:.
 \label{yA-luk PCAdS}
 \end{eqnarray}
 The parameters
$A$, $\Lambda$, $m$ and $q$, written as a function of $y_A$ and
$y_+$, are
\begin{eqnarray}
& & \frac{|\Lambda|}{3A^2} =  \left ( \frac{y_A \: y_+}{y_A+y_+}
\right )^2\,, \nonumber \\
& & mA=(y_A+y_+)^{-1}=qA \,.
 \label{zeros-luk PCAdS}
 \end{eqnarray}
Thus, the mass and the charge of the nonextreme AdS instanton are
necessarily equal, $m=q$, as occurs with its flat
\cite{DGKT,DGGH,HawkRoss} and dS counterparts
\cite{MelMos,MannRoss,OscLem_dS-C}. The demand that $\alpha$ is
real requires that
\begin{eqnarray}
  0< m \leq \frac{1}{4} \sqrt{\frac{3}{3A^2-|\Lambda|}}\:,
 \label{mq-luk-v0}
 \end{eqnarray}
and that
\begin{eqnarray}
 A>\sqrt{|\Lambda|/3}\:.
 \label{Amin}
 \end{eqnarray}
Therefore, as already anticipated, the nonextreme AdS instanton is
available only when (\ref{Amin}) is satisfied.

As we said, the allowed range of $y$ in the Euclidean sector is
$y_A \leq y \leq y_+$. Then, the period of $\tau$, (\ref{Period
tau-yA}), that avoids the conical singularity at both horizons is
\begin{equation}
\beta=\frac{8 \pi \,m A}{\alpha(1-\alpha^2)}\,,
 \label{beta-luk PCAdS}
 \end{equation}
and $T=1/\beta$ is the common temperature of the two horizons.

Using the fact that ${\cal G}(x)=|\Lambda|/(3A^2)-{\cal F}(-x)$
[see (\ref{FG PCAdS})] we can write
\begin{eqnarray}
{\cal G}(x) = 1-x^2 \left ( 1+m A\, x \right )^2 \:,
 \label{G-luk PCAdS}
 \end{eqnarray}
and the roots of ${\cal G}(x)$ we are interested in are the south
and north pole (represented, respectively, as $x_\mathrm{s}$ and
$x_\mathrm{n}$ in Fig. \ref{g3_pc_AdS}),
\begin{eqnarray}
& & x_\mathrm{s} =  \frac{-1+\omega_-}{2mA}<0\,, \:\:\:\:\:\:\:\:
x_\mathrm{n}  = \frac{-1+\omega_+}{2mA}>0\,, \nonumber\\
 & & {\rm
with}\:\:\:\: \omega_{\pm} = \sqrt{1\pm 4mA} \:.
 \label{polos-luk PCAdS}
 \end{eqnarray}
When $m$ and $q$ go to zero we have $x_\mathrm{s}\rightarrow -1$
and $x_\mathrm{n}\rightarrow +1$. This is the reason why we
decided to work in between the roots $x_\mathrm{s}$ and
$x_\mathrm{n}$, instead of working in between the roots
$x'_\mathrm{s}$ and $x'_\mathrm{n}$ also represented in Fig.
\ref{g3_pc_AdS}. Indeed, when $m\rightarrow 0$ and $q\rightarrow
0$ these two last roots disappear, and our instanton has no vacuum
counterpart. Now, the requirement that $\omega_-$ is real demands
that $mA<1/4$ [note that $1/(4A)<(1/4) \sqrt{3/(3A^2-|\Lambda|)}$,
see (\ref{mq-luk-v0})]. If this requirement is not fulfilled then
${\cal G}(x)$ has only two real roots that are represented as
$x'_\mathrm{s}$ and $x_\mathrm{n}$ in Fig. \ref{g3_pc_AdS} and, as
we have just said, in this case the solution has no counterpart in
the $m=0$ and $q=0$ case. Therefore we discard the solutions that
satisfy $\frac{1}{4A}\leq m \leq \frac{1}{4}
\sqrt{\frac{3}{3A^2-|\Lambda|}}$, and hereafter when we refer to
the mass of the nonextreme AdS instanton we will be working in the
range
\begin{eqnarray}
  0< m \leq \frac{1}{4A}\:,
 \label{mq-luk PCAdS}
 \end{eqnarray}

The period of $\phi$, (\ref{Period phi}), that avoids the conical
singularity at the north pole (and leaves one at the south pole
responsible for the presence of the string) is
\begin{equation}
\Delta \phi=\frac{8 \pi\,m A}{\omega_+(\omega^2_+ -1)} <2 \pi\:.
 \label{Period phi-luk PCAdS}
 \end{equation}
When $m$ and $q$ go to zero we have $\Delta \phi \rightarrow 2
\pi$ and the conical singularity disappears.

The topology of the nonextreme AdS instanton is
 $S^2 \times S^2-\{ region \}$, where $S^2 \times S^2$ represents
 $0\leq \tau \leq \beta$,
 $y_A \leq y \leq y_+$, $x_\mathrm{s}\leq x \leq x_\mathrm{n}$,
and $0 \leq \phi \leq \Delta \phi$, but we have to remove the
region, $\{ region \}=\{ \{x, y \}: x_\mathrm{s} \leq x \leq -y_A
\:\: \wedge \:\: y+x=0 \}$. The Lorentzian sector of this
nonextreme instanton describes two charged AdS black holes being
accelerated by the strings, so this instanton describes pair
creation of nonextreme black holes with $m=q$.

\subsection{\label{sec:Cold-inst AdS}The extreme AdS instanton  with $\bm{y_+=y_-}$}
The gravitational field of the extreme AdS instanton  is given by
(\ref{C-metric}) with the requirement that the size of the outer
charged black hole horizon $y_+$ is equal to the size of the inner
charged horizon $y_-$. Let us label this degenerated horizon by
$\rho$: $y_+=y_-\equiv \rho$ and $\rho > y_A$.  In this case, the
function ${\cal F}(y)$ can be written as
\begin{eqnarray}
{\cal F}(y)=\frac{\rho^2-3\gamma}{\rho^4}
 (y-y'_A)(y-y_A)(y-\rho)^2\:,
 \label{F-cold}
 \end{eqnarray}
with
\begin{eqnarray}
\gamma=\frac{3A^2-|\Lambda|}{3A^2}\:, \qquad {\rm and}\:\:\:
A>\sqrt{|\Lambda|/3}\:.
 \label{gamma-PCAdS}
 \end{eqnarray}
Note that, as occurred with the nonextreme AdS instanton, the
extreme AdS instanton is also available only when
$A>\sqrt{|\Lambda|/3}$. The roots $\rho$, $y'_A$ and $y_A$ are
given by
\begin{eqnarray}
& & \rho =\frac{3m}{4q^2A}
 \left ( 1+ \sqrt{1-\frac{8}{9}\frac{q^2}{m^2}} \:\right )
 \:,   \label{zerosy1-cold} \\
& & y'_A =\frac{\gamma \rho}{\rho^2-3\gamma}
 \left ( 1- \sqrt{\frac{\rho^2-2\gamma}{\gamma}} \:\right )
 \:,   \label{zerosy2-cold}\\
& & y_A =\frac{\gamma \rho}{\rho^2-3\gamma}
 \left ( 1+ \sqrt{\frac{\rho^2-2\gamma}{\gamma}} \:\right )
 \:.
 \label{zerosy3-cold}
 \end{eqnarray}
The mass and the charge of the solution are written as a function
of $\rho$ as
\begin{eqnarray}
& & m =\frac{1}{A\rho}
 \left ( 1- \frac{2\gamma}{\rho^2} \right )
 \:,  \nonumber \\
& & q^2 =\frac{1}{A^2\rho^2}
 \left ( 1- \frac{3\gamma}{\rho^2} \right )
 \:,
 \label{mq}
 \end{eqnarray}
and, for a fixed $A$ and $\Lambda$, the ratio $q/m$ is higher than
$1$. The conditions $\rho > y_A$ and $q^2 > 0$ require that, for
the extreme AdS instanton, the allowed range of $\rho$ is
\begin{eqnarray}
\rho>\sqrt{6\gamma} \:.
 \label{range-gamma-cold}
\end{eqnarray}
The value of $y_A$ decreases monotonically with $\rho$ and we have
$\sqrt{\gamma}<y_A<\sqrt{6\gamma}$.
 The mass and the charge of the extreme AdS instanton are also monotonically
decreasing functions of $\rho$, and as we come from $\rho=+\infty$
into $\rho=\sqrt{6\gamma}$ we have
\begin{eqnarray}
& &  0< m< \frac{\sqrt{2}}{3}
  \frac{1}{\sqrt{3A^2-|\Lambda|}}\:, \\
& & 0< q < \frac{1}{2}
  \frac{1}{\sqrt{3A^2-|\Lambda|}}\:,
 \label{mq-cold}
 \end{eqnarray}
so, for a fixed $\Lambda$, as the acceleration parameter $A$ grows
the maximum value of the mass and of the charge decreases
monotonically. For a fixed $\Lambda$ and for a fixed mass below
$\sqrt{2/(9 |\Lambda|)}$, the maximum value of the acceleration is
$\sqrt{2/(27m^2)+|\Lambda|/3}$.

As we have already said, the allowed range of $y$ in the Euclidean
sector is $y_A \leq y < y_+$ and does not include $y=y_+$. Then,
the period of $\tau$, (\ref{Period tau-yA}), that avoids the
conical singularity at the only  horizon of the extreme AdS
instanton is
\begin{equation}
\beta=\frac{2 \pi
\rho^3}{(y_A-\rho)^2\sqrt{\gamma(\rho^2-2\gamma)}}\:,
 \label{beta-cold}
 \end{equation}
and $T=1/\beta$ is the temperature of the acceleration horizon.

In what concerns the angular sector of the extreme AdS instanton,
${\cal G}(x)$ is given by (\ref{FG PCAdS}), and its only real
zeros are the south and north pole (represented, respectively, as
$x_\mathrm{s}$ and $x_\mathrm{n}$ in Fig. \ref{g3_pc_AdS};
$x'_\mathrm{s}$ and $x'_\mathrm{n}$ also represented in this
figure become complex roots in the extreme case),
\begin{eqnarray}
x_\mathrm{s} &=&  -p + \frac{h}{2}-\frac{m}{2q^2A}<0\,, \nonumber \\
 x_\mathrm{n}  &=& +p +
\frac{h}{2}-\frac{m}{2q^2A}>0 \:,
 \label{polos-cold}
 \end{eqnarray}
with
\begin{eqnarray}
p &=& \frac{1}{2} \left (  -\frac{s}{3}+\frac{2m^2}{q^4A^2}
  -\frac{1-12 q^2A^2}{3s q^4A^4} -\frac{4}{3q^2A^2}
  + n\right )^{1/2} \:,  \nonumber \\
n &=&  \frac{-m^3+mq^2}{2hq^6 A^3} \:,  \nonumber \\
h &=& \sqrt{\frac{s}{3}+\frac{m^2}{q^4A^2}+\frac{1-12 q^2A^2}{3s q^4A^4}
-\frac{2}{3q^2A^2}} \:, \nonumber \\
 s &=& \frac{1}{2^{1/3} q^2A^2}\left (
\lambda-\sqrt{\lambda^2-4(1-12 q^2A^2)^3} \right )^{1/3} \:,
\nonumber \\
\lambda &=& 2 - 108 m^2A^2 + 72 q^2A^2\:,
  \label{acess-zeros-ang}
 \end{eqnarray}
where $m$ and $q$ are fixed by (\ref{mq}), for a given $A$,
$\Lambda$ and $\rho$. When $m\rightarrow 0$ and $q\rightarrow 0$
we have $x_\mathrm{s}\rightarrow -1$ and $x_\mathrm{n}\rightarrow
+1$. The period of $\phi$ that avoids the conical singularity at
the north pole (and leaves one at the south pole responsible for
the presence of the strings) is given by (\ref{Period phi}) with
$x_\mathrm{n}$ defined in (\ref{polos-cold}).

The topology of the extreme AdS instanton is ${\mathbb{R}}^2
\times S^2-\{ region \}$, where ${\mathbb{R}}^2 \times S^2$
represents $0 \leq \tau \leq \beta$, $y_A \leq y < y_+$,
 $x_\mathrm{s}\leq x \leq x_\mathrm{n}$,
and $0 \leq \phi \leq \Delta \phi$, but we have to remove the
region, $\{ region \}=\{ \{x, y \}: x_\mathrm{s} \leq x \leq -y_A
\:\: \wedge \:\: y+x=0 \}$.  Since $y=y_+=\rho$ is at an infinite
proper distance, the surface $y=y_+=\rho$ is an internal infinity
boundary. The Lorentzian sector of this extreme case describes two
extreme ($y_+=y_-$) charged AdS black holes being accelerated by
the strings, and the extreme AdS instanton describes the pair
creation of these extreme black holes.

\subsection{\label{sec:submaximal-inst AdS}The nonextreme AdS instanton
with $\bm{m\neq q}$}

The AdS C-metric instantons studied in the two last subsections,
namely the nonextreme instantons with $m=q$ and the extreme
instantons with $y_+=y_-$,  are saddle point solutions free of
conical singularities both in the $y_+$ and $y_A$ horizons. The
corresponding black holes may then nucleate in the AdS background
when a cosmic string breaks, and we will compute their pair
creation rates in Sec. \ref{sec:Calc-I AdS}. However, these
particular black holes are not the only ones that can be pair
created. Indeed, it has been shown in
\cite{WuSubMaxBoussoHawkSubMax} that Euclidean solutions with
conical singularities may also be used as saddle points for the
pair creation process. These nonextreme instantons have $m\neq q$
and describe pair creation of nonextreme black holes with $m\neq
q$.

\begin{figure}[h]
\includegraphics[height=6cm]{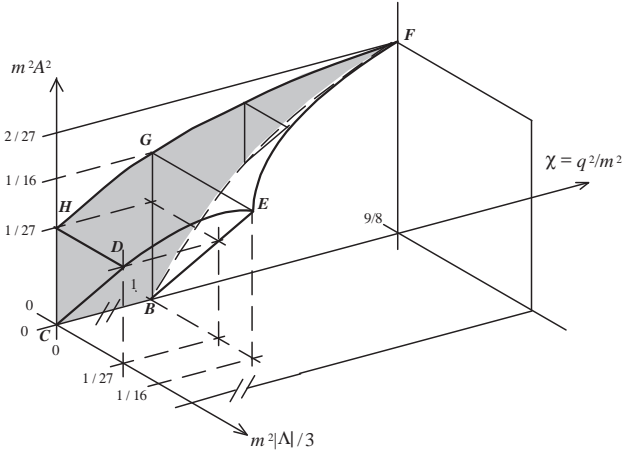}
   \caption{\label{Fig-range pc AdS}
Allowed ranges of the parameters $\Lambda, A, m, \chi\equiv
q^2/m^2$ for which one has a solution representing a pair of
accelerated black holes. The planar surface whose frontier is the
triangle $BEG$ represents the nonextreme AdS instanton with $m=q$.
The curved surface delimitated by the closed line $BEF$ represents
the extreme AdS instanton with $y_+=y_-$ and satisfies $m^2A^2 =
\sigma(\chi,\gamma_+)+m^2|\Lambda|/3$. The curved surface whose
frontier is $DEFGH$ satisfies $m^2A^2=\sigma(\chi,\gamma_-)$ [see
(\ref{rangeG<0})]. The plane surface with boundary given by $BCDE$
satisfies $A=\sqrt{|\Lambda|/3}$. Neutral AdS instantons ($q=0$)
are those that belong to the planar surface with the triangle
boundary $CDH$. The nonextreme AdS instantons with $m\neq q$ are
those whose parameters are in the volume with boundary defined by
$BCDE$, $BEF$, $CBFGH$, $CDH$ and $DEFGH$.
 }
\end{figure}

In what follows we will find the range of parameters for which one
has nonextreme black holes with conical singularities, i.e., with
$m\neq q$. First, when $m\neq 0$ and $q\neq 0$, we require that
$x$ belongs to the interval $[x_\mathrm{s},x_\mathrm{n}]$
(sketched in Fig. \ref{g3_pc_AdS}) for which the charged solutions
are in the same sector of the $m=0$ and $q=0$ solutions. Defining
\begin{eqnarray}
& & \chi \equiv \frac{q^2}{m^2}\:,  \;\;\; 0<\chi\leq \frac{9}{8}
\:, \;\;\;\;\;\gamma_{\pm} \equiv
1 \pm \sqrt{1-\frac{8}{9}\chi} \:, \nonumber \\
 & &
\sigma(\chi,\gamma_{\pm})=
\frac{(4\chi)^2(3\gamma_{\pm})^2-8\chi(3\gamma_{\pm})^3+\chi(3\gamma_{\pm})^4}
{(4\chi)^4} \:,  \nonumber \\
 \label{beta dS}
\end{eqnarray}
the above requirement is fulfilled by the parameter range
\begin{equation}
m^2A^2<\sigma(\chi,\gamma_-)\:,
 \label{rangeG<0}
 \end{equation}
for which ${\cal G}(x=x_0)<0$, with
$x_0=-\frac{3\gamma_-}{4\chi}\frac{1}{mA}$ being the less negative
$x$ where the derivative of ${\cal G}(x)$ vanishes. Second, in
order to insure that one has a nonextreme solution we must require
that
\begin{equation}
m^2A^2 > \sigma(\chi,\gamma_+)+m^2|\Lambda|/3\:,
 \label{rangeF>0}
 \end{equation}
 for which ${\cal F}(y=y_0)<0$,
with $y_0=\frac{3\gamma_+}{4\chi}\frac{1}{mA}$ being the point in
between $y_+$ and $y_-$ where the derivative of ${\cal F}(y)$
vanishes. We have $\sigma(\chi,\gamma_-)>\sigma(\chi,\gamma_+)$
except at $\chi=9/8$ where these two functions are equal;
$\sigma(\chi,\gamma_-)$ is always positive; and
$\sigma(\chi,\gamma_+)<0$ for $0<\chi<1$ and
$\sigma(\chi,\gamma_+)>0$ for $1<\chi\leq 9/8$. The nonextreme
black holes with conical singularities are those that satisfy
(\ref{rangeG<0}), (\ref{rangeF>0}) and $A>\sqrt{|\Lambda|/3}$. The
range of parameters of these nonextreme black holes with $m\neq q$
are sketched in Fig. \ref{Fig-range pc AdS}.

In order to compute the pair creation rate of the nonextreme black
holes with $m\neq q$, we will need the relation between the
parameters $A$, $\Lambda$, $m$, $q$, and the horizons $y_A$, $y_+$
and $y_-$. In general, for a nonextreme solution with horizons
$y_A<y_+<y_-$, one has
\begin{eqnarray}
{\cal F}(y)= -\frac{1}{d}(y-y_A)(y-y_+)(y-y_-)(ay+b) \:,
 \label{F-nonext-sub-max pcAdS}
 \end{eqnarray}
with
\begin{eqnarray}
d\!\!&=&\!\! y_A y_+ y_- (y_A +y_+ +y_-) +(y_A y_+ +y_A y_- +y_+
y_-)^2 \nonumber \\
 a\!\!&=&\!\!
 \left ( y_A y_+ +y_A y_- +y_+y_- \right )  \nonumber \\
 b\!\!&=&\!\! y_A y_+y_-\:.
 \label{F-nonext-sub-max aux pcAdS}
 \end{eqnarray}
The parameters $A$, $\Lambda$, $m$ and $q$ can be expressed as a
function of $y_A$, $y_+$ and $y_-$ by
\begin{eqnarray}
\frac{|\Lambda|}{3A^2}\!\!&=&\!\! 1-d^{-2}(y_A y_+y_-)^2 \nonumber \\
 q^2A^2\!\!&=&\!\! d^{-1}(y_A y_+ +y_A y_-
+y_+ y_-) \nonumber \\
 mA\!\!&=&\!\! (2\,\sigma)^{-1}(y_A +y_+)(y_A +y_-)(y_+ +y_-) \nonumber \\
 \sigma\!\!&=&\!\! y_A^2 y_+y_- +y_A y_+^2 y_- +y_A y_+ y_-^2+ (y_A y_+)^2 \nonumber \\
  & & +(y_A y_-)^2 +(y_+ y_-)^2 \:.
 \label{relation parameters nonext-sub-max pcAdS}
 \end{eqnarray}
The allowed values of parameters $m$ and $q$ are those contained
in the interior region sketched in Fig. \ref{Fig-range pc AdS}.

The topology of these nonextreme AdS instantons with $m\neq q$ is
 $S^2 \times S^2-\{ region \}$, where $S^2 \times S^2$ represents
 $0\leq \tau \leq \beta$,
 $y_A \leq y \leq y_+$, $x_\mathrm{s}\leq x \leq x_\mathrm{n}$,
and $0 \leq \phi \leq \Delta \phi$, but we have to remove the
region, $\{ region \}=\{ \{x, y \}: x_\mathrm{s} \leq x \leq -y_A
\:\: \wedge \:\: y+x=0 \}$. These instantons describe the pair
creation of nonextreme black holes with $m \neq q$.

\subsection{\label{sec:background string AdS}Initial system: AdS background with a string}

So far, we have described the solution that represents a pair of
black holes accelerated by the strings tension. This solution
describes the evolution of the black hole pair after its creation.
Now, we want to find a solution that represents a string in a AdS
background. This solution will describe the initial system, before
the breaking of the cosmic string that leads to the formation of
the black hole pair. In order to achieve our aim we note that at
spatial infinity the gravitational field of the Euclidean AdS
C-metric reduces to
\begin{eqnarray}
\!\!\!\!\!\!\!\!\! ds^2 \!\! &=& \!\!
\frac{1}{[A_0(x+y)]^2}{\biggl [}
 -\left ( \frac{|\Lambda|}{3A_0^{\,2}}-1+y^2 \right )dt^2
 \nonumber\\
 & & +  \frac{dy^2}{\frac{|\Lambda|}{3A_0^{\,2}}-1+y^2}
+\frac{dx^2}{1-x^2}+ (1-x^2)d\phi_0^{\,2} {\biggr ]},
 \label{backg metric}
\end{eqnarray}
and the Maxwell field goes to zero. $A_0$ is a constant that
represents a freedom in the choice of coordinates, and $-1 \leq x
\leq 1$. We want that this metric also describes the solution
before the creation of the black hole pair, i.e., we demand that
it describes a string with its conical deficit in an AdS
background. Now, if we want to maintain the intrinsic properties
of the string during the process we must impose that its mass
density and thus its conical deficit remains constant. After the
pair creation we already know that the conical deficit is given by
(\ref{conic-sing}). Hence, the requirement that the background
solution describes an AdS spacetime with a conical deficit angle
given exactly by (\ref{conic-sing}) leads us to impose that in
(\ref{backg metric}) one has
\begin{eqnarray}
\Delta \phi_0=2\pi-\delta=2\pi \frac{{\cal
G}'(x_\mathrm{s})}{|{\cal G}'(x_\mathrm{n})|} \,.
 \label{delta phi0}
 \end{eqnarray}
The arbitrary parameter $A_0$ can be fixed as a function of $A$ by
imposing a matching between (\ref{C-metric}) and
 (\ref{backg metric}) at large spatial distances. We will do this matching
in the following section.

\section{\label{sec:Calc-I AdS}Calculation of the black hole
pair creation rates}
The black hole pair creation rate is given by the path integral
 \begin{eqnarray}
\Gamma(g_{ij},A_{i})=\int d[g_{\mu\nu}]d[A_{\mu}]
 e^{-\left [ I(g_{\mu\nu},A_{\mu})-I_0(g^0_{\mu\nu},A^0_{\mu})\right ]}\:,
 \label{path integral}
 \end{eqnarray}
where $g_{ij}$ and $A_{i}$ are the induced metric and
electromagnetic potential on the boundary
 $\partial {\cal M}$ of a compact manifold ${\cal M}$,
 $d[g_{\mu\nu}]$ is a measure on the space of the metrics
$g_{\mu\nu}$ and $d[A_{\mu}]$ is a measure on the space of Maxwell
field $A_{\mu}$, and $I(g_{\mu\nu},A_{\mu})$ is the Euclidean
action of the instanton that mediates the process. In our case
this action is the Einstein-Maxwell action with a negative
cosmological constant $\Lambda$. The path integral is over all
compact metrics and potentials on manifolds ${\cal M}$ with
boundary $\partial {\cal M}$, which agree with the boundary data
on $\partial {\cal M}$. $I_0(g^0_{\mu\nu},A^0_{\mu})$ is the
action for the background reference spacetime, the AdS background
with the string, specified by $g^0_{\mu\nu}$ and $A^0_{\mu}$ (see
Sec. \ref{sec:background string AdS}). Its presence is required
because it describes the initial system. Moreover, the geometry of
the final system with the black hole pair is noncompact, that is
$I(g_{\mu\nu},A_{\mu})$ diverges. However, the physical action
$I(g_{\mu\nu},A_{\mu})-I_0(g^0_{\mu\nu},A^0_{\mu})$ is finite for
fields $g_{\mu\nu}$, $A_{\mu}$ that approach asymptotically
$g^0_{\mu\nu}$, $A^0_{\mu}$ in an appropriate way \cite{HawkHor}.
Specifically, one fixes a boundary near infinity,
$\Sigma^{\infty}$, and demands that $g_{\mu\nu}$, $A_{\mu}$ and
$g^0_{\mu\nu}$, $A^0_{\mu}$ induce fields on this boundary  that
agree to sufficient order, so that their difference does not
contribute to the action in the limit that $\Sigma^{\infty}$ goes
to infinity \cite{HawkHor}.

In the semiclassical instanton approximation, the dominant
contribution to the path integral (\ref{path integral}) comes from
metrics and Maxwell fields which are near the solutions
(instantons) that extremalize the Euclidean action and satisfy the
boundary conditions. In this approximation, the pair creation rate
of AdS black holes is then given by
\begin{eqnarray}
\Gamma \sim \eta \,e^{-I_{\rm inst}} \:,
 \label{PC-rate}
 \end{eqnarray}
where $I_{\rm inst}\equiv I-I_0$  includes already the
contribution from the background reference spacetime. $\eta$ is
the one-loop prefactor that accounts for the fluctuations in the
gravitational and matter fields, and its evaluation will not be
considered in this paper (see
\cite{GaratinniOneLoop,VolkovWipf,GrossPerryYaffe,OneLoop} for a
treatment of this factor in some backgrounds).

Hawking and Horowitz \cite{HawkHor} (see also Brown and York
\cite{BrownYork}) have shown that the Euclidean action of the
instanton that mediates pair creation of nonextreme black holes
can be written in the form
 \begin{eqnarray}
I_{\rm inst}=\beta H-\frac{1}{4}
 \left(\Delta {\cal A}_{\rm ac}+{\cal A}_{\rm bh}\right )\:,
 \label{I-Non-Ext}
 \end{eqnarray}
where $\Delta {\cal A}_{\rm ac}$ is the difference in area of the
acceleration horizon between the AdS C-metric and the background,
and ${\cal A}_{\rm bh}$ is the area of the black hole horizon
present in the instanton. The Euclidean action of the instanton
that describes pair creation of extreme black holes can be written
as
\begin{eqnarray}
I_{\rm inst}=\beta H-\frac{1}{4}\Delta {\cal A}_{\rm ac}\:.
 \label{I-Ext}
 \end{eqnarray}
In these relations, $\beta$ is the period of the Euclidean time,
and $H$ represents the Hamiltonian of the system which, for static
solutions, is given by \cite{HawkHor,BrownYork}
\begin{eqnarray}
H\!\!&=&\!\!\int_{\Sigma_t}\!\!d^3x\sqrt{h}\,N{\cal
H}-\frac{1}{8\pi}\int_{\Sigma_t^{\infty}}\!\!\!d^2x\sqrt{\sigma}
\left ( N\, {}^2\!K-N_0\, {}^2\!K_0 \right ). \nonumber \\
 & &
 \label{hamiltonian}
 \end{eqnarray}
 The boundary $\partial M$ consists of an initial and final
spacelike surface, $\Sigma_t$, of constant $t$ with unit normal
$u^{\mu}$ ($u\cdot u=-1$) and with intrinsic metric
$h_{\mu\nu}=g_{\mu\nu}+u_{\mu}u_{\nu}$ plus a timelike surface
near infinity, $\Sigma^{\infty}$, with unit normal $n^{\mu}$
($n\cdot n=1$, and $u\cdot n=0$) and with intrinsic metric
$\sigma_{\mu\nu}=h_{\mu\nu}-n_{\mu}n_{\nu}$. This surface
$\Sigma^{\infty}$ needs not to be at infinity. It can also be at a
black hole horizon or at an internal infinity, but we will
generally label it by $\Sigma^{\infty}$. This surface
$\Sigma^{\infty}$ is foliated by a family of 2-surfaces
$\Sigma_t^{\infty}$ that result from the intersection between
$\Sigma^{\infty}$ and $\Sigma_t$. In (\ref{hamiltonian}), $N$ and
$N_0$ are the lapse functions of the system with the pair and of
the background, respectively, and $N=N_0$ in the boundary near
infinity, $\Sigma^{\infty}$.
 ${\cal H}$ is the hamiltonian constraint that contains
contributions from the gravitational and Maxwell fields, and
vanishes for solutions of the equations of motion. Then, the
Hamiltonian is simply given by the boundary surface term. In this
surface term, ${}^2\!K$ represents the trace of the extrinsic
curvature of the surface imbedded in the AdS C-metric, and
${}^2\!K_0$ is the extrinsic curvature of the surface imbedded in
the background spacetime.

We will now verify that the boundary surface term in the
Hamiltonian (\ref{hamiltonian}) is also zero, and thus the
Hamiltonian makes no contribution to (\ref{I-Non-Ext}) and
(\ref{I-Ext}). We follow the technical procedure applied by
Hawking, Horowitz and Ross \cite{HawHorRoss} in the Ernst solution
and by  Hawking and Ross \cite{HawkRoss-string} in the flat
C-metric. As we said above, one will require that the intrinsic
metric $ds^2_{(\Sigma_t^{\infty})}$ on the boundary
$\Sigma_t^{\infty}$ as embedded in the AdS C-metric agrees (to
sufficient order) with the intrinsic metric on the boundary
$\Sigma_t^{\infty}$ as embedded in the background spacetime, in
order to be sure that one is taking the same near infinity
boundary in the evaluation of the quantities in the two
spacetimes. In the AdS C-metric one takes this boundary to be at
$x+y=\varepsilon_c$, where $\varepsilon_c \ll 1$. The background
reference spacetime is described by (\ref{backg metric}),
subjected to $-1 \leq x \leq 1$, $y \geq -x$ and (\ref{delta
phi0}). In this background spacetime we take the boundary
$\Sigma_t^{\infty}$ to be at $x+y=\varepsilon_0$, where
$\varepsilon_0 \ll 1$.

We are now interested in writing the intrinsic metric in the
boundary $\Sigma_t^{\infty}$ (the 2-surface $t=$ const and
$x+y=\varepsilon$). In the AdS C-metric
 (\ref{C-metric}) one performs the coordinate transformation
\begin{eqnarray}
\phi=\frac{2}{|{\cal G}'(x_\mathrm{n})|}\,\tilde{\phi}\,, \qquad
t=\frac{2}{{\cal F}'(y_A)}\,\tilde{t}\,
 \label{redef phi t}
 \end{eqnarray}
in order that $\Delta \tilde{\phi}=2\pi$, and the analytic
continuation of $\tilde{t}$ has period $2\pi$, i.e., $\Delta
\tilde{\tau}=2\pi$. Furthermore, one takes
\begin{eqnarray}
x=x_\mathrm{s}+\varepsilon_c \chi \,, \qquad
y=-x_\mathrm{s}+\varepsilon_c (1-\chi)\,
 \label{def varepsilon chi C}
 \end{eqnarray}
where $0 \leq \chi \leq 1$. By making the evaluations up to second
order in $\varepsilon_c$ (since higher order terms will not
contribute to the Hamiltonian in the final limit $\varepsilon_c
\rightarrow 0$), the intrinsic metric on the boundary
$\Sigma_t^{\infty}$ is then
\begin{eqnarray}
 \!\!\!\!\!\!\!\!\!\!\!\!ds^2_{(\Sigma_t^{\infty})} &\sim&
\frac{2}{A^2\varepsilon_c {\cal G}'(x_\mathrm{s}) }
 {\biggl [}
\frac{d \chi^2}{2\chi} \nonumber\\
& &+ \left |
 \frac{{\cal G}'(x_\mathrm{s})}{{\cal G}'(x_\mathrm{n})}
 \right |^2 \!\!\!\left ( 2\chi+ \varepsilon_c
 \frac{{\cal G}''(x_\mathrm{s})}{{\cal G}'(x_\mathrm{s})}\chi^2 \right )
 d \tilde{\phi}^2  {\biggr ]}.
 \label{metric boundary C}
 \end{eqnarray}
Analogously, for the background spacetime (\ref{backg metric}) one
sets
\begin{eqnarray}
x=-1+\varepsilon_0 \chi\, , \qquad y=1+\varepsilon_0 (1-\chi)\,
 \label{def varepsilon chi bg}
 \end{eqnarray}
and the intrinsic metric on the boundary $\Sigma_t^{\infty}$
yields
\begin{eqnarray}
ds^2_{(\Sigma_t^{\infty})}\sim  \frac{1}{A_0^{\,2}\varepsilon_0}
 \left [ \frac{d \chi^2}{2\chi}+ \left ( 2\chi- \varepsilon_0
 \chi^2 \right )
 d\phi_0^{\,2}  \right ]
 \label{metric boundary bg}
 \end{eqnarray}
These two intrinsic metrics on the boundary will agree (up to
second order in $\varepsilon$) as long as we take the period of
$\phi_0$ to be given by (\ref{delta phi0}) and the following
matching conditions are satisfied,
\begin{eqnarray}
\varepsilon_0=- \frac{{\cal G}''(x_\mathrm{s})}{{\cal
G}'(x_\mathrm{s})}\,\varepsilon_c \,,
 \label{match condition epsilon}
 \end{eqnarray}
\begin{eqnarray}
A_0^{\,2}=- \frac{[{\cal G}'(x_\mathrm{s})]^2}{2{\cal
G}''(x_\mathrm{s})}\,A^2  \,.
 \label{match condition A}
 \end{eqnarray}
Note that the Maxwell fields of the two solutions agree trivially
at the near infinity boundary $\Sigma_t^{\infty}$.

In what concerns the lapse function of the AdS C-metric, we
evaluate it with respect to the time coordinate $\tilde{t}$
defined in (\ref{redef phi t}) and, using $[A(x+y)]^{-2}{\cal
F}dt^2=N^2 d\tilde{t}^{\,2}$, we find
\begin{eqnarray}
N\!\!&\sim&\!\!
\sqrt{\frac{|\Lambda|}{3}}\frac{2}{A^2\varepsilon_c {\cal F}'(y_A)
}
 \left ( 1+\frac{1}{2}\frac{3}{|\Lambda|}\left (1-\chi \right )
 A^2\varepsilon_c \,{\cal G}'(x_\mathrm{s})
  \right ) . \nonumber \\
 & &
 \label{lapse function C}
 \end{eqnarray}
Analogously, an evaluation with respect to the time coordinate $t$
defined in (\ref{backg metric}) yields
\begin{eqnarray}
N_0\sim \frac{{\cal G}'(x_\mathrm{s})}{{\cal F}'(y_A)}
\sqrt{\frac{|\Lambda|}{3}}\frac{1}{A_0^{\,2}\varepsilon_0}
 \left ( 1+\frac{3}{|\Lambda|}\left (1-\chi \right )
 A_0^{\,2}\varepsilon_0
  \right ) .
 \label{lapse function bg}
 \end{eqnarray}
Note that these two lapse functions are also matched by the
conditions (\ref{match condition epsilon}) and (\ref{match
condition A}).

The extrinsic curvature to $\Sigma_t^{\infty}$ as embedded in
$\Sigma_t$ is ${}^2\!K_{\mu\nu}=\sigma_{\mu}^{\:\:\:\alpha}
h_{\alpha}^{\:\:\:\beta}\nabla_{\beta}n_{\nu}$ (where
$\nabla_{\beta}$ represents the covariant derivative with respect
 to $g_{\mu\nu}$), and the trace of the extrinsic curvature is
${}^2\!K=g^{\mu\nu}\:{}^2\!K_{\mu\nu}=A\sqrt{{\cal F}(y)}$. The
extrinsic curvature of the boundary embedded in the AdS C-metric
is then
\begin{eqnarray}
{}^2\!K \sim \sqrt{\frac{|\Lambda|}{3}}
 \left ( 1+\frac{1}{2}\frac{3}{|\Lambda|}\left (1-\chi \right )
 A^2\varepsilon_c \,{\cal G}'(x_\mathrm{s})
  \right )\,,
 \label{extrinsic curvat C}
 \end{eqnarray}
while the extrinsic curvature of the boundary embedded in the
background reference spacetime is
\begin{eqnarray}
{}^2\!K_0\sim \sqrt{\frac{|\Lambda|}{3}}
 \left ( 1+\frac{3}{|\Lambda|}\left (1-\chi \right )
 A_0^{\,2}\varepsilon_0 \right ) .
 \label{extrinsic curvat bg}
 \end{eqnarray}

We are now in position to compute the contribution from the
surface boundary term in (\ref{hamiltonian}). The evaluation in
the AdS C-metric yields
\begin{eqnarray}
\!\!\!\!\!\!\!\! \int_{\Sigma_t^{\infty}}\!\!\!d\chi d\tilde{\phi}
\sqrt{\sigma}\,
 N\,{}^2\!K  &\sim& \frac{8\pi}{|{\cal G}'(x_\mathrm{n})|{\cal F}'(y_A)}
 \frac{|\Lambda|}{3} \frac{1}{(A^2\varepsilon_c)^2} \nonumber\\
 & & \times
 \left ( 1-\frac{1}{2}\frac{3}{|\Lambda|}
 A^2\varepsilon_c \,{\cal G}'(x_\mathrm{s})
  \right ) ,
 \label{int NK C}
 \end{eqnarray}
while for the background reference spacetime we have
\begin{eqnarray}
\!\!\!\!\!\!\!\!
\int_{\Sigma_t^{\infty}}\!\!\!\sqrt{\sigma_0}d\chi d\phi_0
 N_0\,{}^2\!K_0 &\sim&
 \frac{8\pi \,[{\cal G}'(x_\mathrm{s})]^2}{|{\cal G}'(x_\mathrm{n})|{\cal F}'(y_A)}
 \frac{|\Lambda|}{3}  \frac{1}{(2A_0^{\,2}\varepsilon_0)^2} \nonumber\\
 & & \times \left ( 1-\frac{3}{|\Lambda|}
 A_0^{\,2}\varepsilon_0 \right ) .
 \label{int NK bg}
 \end{eqnarray}
From the matching conditions (\ref{match condition epsilon}) and
(\ref{match condition A}) we conclude that these two boundary
terms are equal. Hence, the surface term and the Hamiltonian
(\ref{hamiltonian}) vanish. The Euclidean action (\ref{I-Non-Ext})
of the nonextreme AdS instanton that mediates pair creation of
nonextreme black holes is then simply
\begin{eqnarray}
I_{\rm nonext}=-\frac{1}{4}
 \left(\Delta {\cal A}_{\rm ac}+{\cal A}_{\rm bh}\right )\:,
 \label{I-Non-Ext2}
 \end{eqnarray}
while the Euclidean action (\ref{I-Non-Ext}) of the extreme AdS
instanton that mediates pair creation of nonextreme black holes is
just given by
\begin{eqnarray}
I_{\rm ext}=-\frac{1}{4}\Delta {\cal A}_{\rm ac}\:.
 \label{I-Ext2}
 \end{eqnarray}
Thus as occurs in the Ernst case, in the flat C-metric case, in
the de Sitter case and in the dS C-metric case, the pair creation
of nonextreme black holes is enhanced relative to the pair
creation of extreme black holes by a factor of $e^{{\cal A}_{\rm
bh}}$.

In the next two subsections we will explicitly compute
(\ref{I-Non-Ext2}) using the results of Sec.
\ref{sec:Lukewarm-inst AdS}, and (\ref{I-Ext2}) using the results
of Sec. \ref{sec:Cold-inst AdS}.
 In Sec. \ref{sec:submaximal-rate AdS} we analyze the pair creation
rate of black holes discussed in Sec.
 \ref{sec:submaximal-inst AdS}.
 Remark that the only horizons that contribute with their
areas to (\ref{I-Non-Ext2}) and (\ref{I-Ext2})  are those that
belong to the instanton responsible for the pair creation, i.e,
only those horizons that are in the Euclidean sector of the
instanton will make a contribution.

The domain of validity of our results is the  particle limit,
$mA\ll 1$, for which the  radius of the black hole, $r_+ \sim m$,
is much smaller than the typical distance between the black holes
at the creation moment, $\ell \sim 1/A$ (this value follows from
the Rindler motion $x^2-t^2=1/A^2$ that describes the uniformly
accelerated motion of the black holes).

\subsection{\label{sec:Lukewarm-rate AdS}Pair creation rate  in
the nonextreme case with $\bm{m=q}$}

In the nonextreme case, the instanton has two horizons in its
Euclidean section, namely the acceleration horizon at $y=y_A$ and
the black hole horizon at $y=y_+$ (the horizons $y'_A$ and $y_-$
do not belong to the instanton, see Fig \ref{g3_pc_AdS}). The
black hole horizon covers the whole range of the angular
coordinate $x$,
 $x_\mathrm{s} \leq x\leq x_\mathrm{n}$, and its area is
\begin{eqnarray}
\cal{A}_{\rm bh} \!&=&\!  \int_{y=y_+} \!\!\!\!\!\!
\sqrt{g_{xx}g_{\phi\phi}}\: dx \,d\phi  \nonumber \\
&=& \frac{1}{A^2}\int_{\Delta \phi}\!\! d\phi
  \int_{x_\mathrm{s}}^{x_\mathrm{n}} \!\! \frac{dx}{(x+y_+)^2}    \nonumber \\
&=&
 \frac{4\pi} {A^2 |{\cal G}'(x_\mathrm{n})|}\,
 \frac{ x_\mathrm{n}-x_\mathrm{s} }
   {(x_\mathrm{n}+y_+)(x_\mathrm{s}+y_+)}\,,
\label{area BH-luk PCAdS}
 \end{eqnarray}
where $y_+$ is given in (\ref{yA-luk PCAdS}), $x_\mathrm{s}$ and
$x_\mathrm{n}$ are defined by (\ref{polos-luk PCAdS}), and
 $\Delta \phi$ is given by (\ref{Period phi-luk PCAdS}).

  The acceleration horizon, $y_A$ defined in (\ref{yA-luk PCAdS}), of
the nonextreme AdS instanton covers the angular range $-y_A \leq
x\leq x_\mathrm{n}$ (i.e., it is not present in the vicinity of
the south pole), and is noncompact, i.e., its area is infinite. We
have to deal appropriately with this infinity. In order to do so,
we first introduce a boundary at $x=-y_A+\varepsilon_c$
($\varepsilon_c \ll 1$), and compute the area inside of this
boundary, which yields
\begin{eqnarray}
{\cal A}^{c}_{\rm ac} \!&=&\!  \int_{y=y_A} \!\!\!\!\!\!
\sqrt{g_{xx}g_{\phi\phi}}\: dx \,d\phi\nonumber \\
&=& \frac{1}{A^2}\int_{\Delta \phi}\!\! d\phi
  \int_{-y_A+\varepsilon_c}^{x_\mathrm{n}}  \frac{dx}{(x+y_A)^2}    \nonumber \\
&=& - \frac{4\pi} {A^2 |{\cal G}'(x_\mathrm{n})|}\,
 \left ( \frac{1}{ x_\mathrm{n}+y_A}
  -\frac{1}{\varepsilon_c} \right )\,,
\label{area Ac-luk PCAdS}
 \end{eqnarray}
When we let $\varepsilon_c \rightarrow 0$, the term
$1/\varepsilon_c$ diverges, and the acceleration horizon has an
infinite area. This area is renormalized with respect to the area
of the acceleration horizon of the background reference spacetime
(\ref{backg metric}). This background reference spacetime has  an
acceleration horizon at $y=\sqrt{1-|\Lambda|/(3A_0^{\,2})}$ which
is the direct counterpart of the above acceleration horizon of the
nonextreme AdS instanton, and it covers the angular range
$-\sqrt{1-|\Lambda|/(3A_0^{\,2})} \leq x\leq 1$. The area inside a
boundary at $x=-\sqrt{1-|\Lambda|/(3A_0^{\,2})}+\varepsilon_0$ is
\begin{eqnarray}
{\cal A}^{0}_{\rm ac} \!&=&\!
  \int_{y=\sqrt{1-\frac{|\Lambda|}{3A_0^{\,2}} }}
\sqrt{g_{xx}g_{\phi_0 \phi_0}}\: dx \,d\phi_0  \nonumber \\
\!&=&\! \frac{1}{A_0^{\,2} }\int_{\Delta \phi_0}\!\! d\phi_0
  \int_{-\sqrt{1-\frac{|\Lambda|}{3A_0^{\,2}} }+\varepsilon_0}^{1}
  \frac{dx}{\left ( x+\sqrt{1-\frac{|\Lambda|}{3A_0^{\,2}}}\right )^2}
   \nonumber \\
\!&=&\! - \frac{2\pi} {A_0^{\,2}} \frac{{\cal
G}'(x_\mathrm{s})}{|{\cal G}'(x_\mathrm{n})|}\,
 \left ( \frac{1}{1+\sqrt{1-\frac{|\Lambda|}{3A_0^{\,2}}} }
  -\frac{1}{\varepsilon_0} \right )   \,,
  \label{area Ac-luk-bg}
 \end{eqnarray}
where in the last step we have replaced $\Delta \phi_0$ by
(\ref{delta phi0}). When $\varepsilon_0 \rightarrow 0$, the term
$1/\varepsilon_0$ diverges, and thus the area of this background
acceleration horizon is also infinite. Now, we have found that the
intrinsic metrics at the near infinity boundary match together if
(\ref{match condition epsilon}) and (\ref{match condition A}) are
satisfied. Our next task is to verify that these matching
conditions between ($\varepsilon_c, A$) and ($\varepsilon_0, A_0$)
are such that the divergent terms in (\ref{area Ac-luk PCAdS}) and
(\ref{area Ac-luk-bg}) cancel each other. It is straightforward to
show that $\Delta {\cal A}_{\rm ac}={\cal A}^{c}_{\rm ac}-{\cal
A}^{0}_{\rm ac}$ yields a finite value if
\begin{eqnarray}
 2A_0^{\,2}\varepsilon_0= A^2\varepsilon_c {\cal
 G}'(x_\mathrm{s})\,.
  \label{match condition}
 \end{eqnarray}
Thus, the matching conditions (\ref{match condition epsilon}) and
(\ref{match condition A}) satisfy condition
 (\ref{match condition}), i.e., they indeed eliminate the divergencies in
$\Delta {\cal A}_{\rm ac}$. It is worthy to remark that with the
choices (\ref{match condition epsilon}) and (\ref{match condition
A}) the proper lengths of the boundaries $x=-y_A+\varepsilon_c$
and $x=-1+\varepsilon_0$, (given, respectively, by $l_c=\int
\sqrt{g_{\phi\phi}}d\phi$ and $l_0=\int
\sqrt{g_{\phi_0\phi_0}}d\phi_0$) do not match. This is in contrast
with the flat case \cite{HawkRoss-string}, in which the choice of
the matching parameters that avoids the infinities in $\Delta
{\cal A}_{\rm ac}$, also leads to $l_c=l_0$. In the AdS case, our
main goal was to remove the infinities in $\Delta {\cal A}_{\rm
ac}$. We have achieved this aim by comparing the appropriate
acceleration horizons in the instanton and in the reference
background. The fact that the matching relations then lead to $l_c
\neq l_0$ is not a problem at all. Replacing  (\ref{match
condition epsilon}) and (\ref{match condition A}) in (\ref{area
Ac-luk-bg}) yields for $\Delta {\cal A}_{\rm ac}={\cal A}^{c}_{\rm
ac}-{\cal A}^{0}_{\rm ac}$ the result
\begin{eqnarray}
& &\!\!\!\!\!\!\!\!\!\!\!\!\!\!\!\!\! \Delta {\cal A}_{\rm ac} =
   -\frac{4\pi} {A^2 |{\cal G}'(x_\mathrm{n})|}\,  \nonumber\\
 & & \!\!\!\!\!\!\!\!\! \times
 {\biggl (} \frac{1}{ x_\mathrm{n}+y_A}
 + \frac{1}{ 1+\sqrt{1-\frac{|\Lambda|}{3A^2}
\frac{2|{\cal G}''(x_\mathrm{s})|}{[{\cal G}'(x_\mathrm{s})]^2} }}
 \frac{ {\cal G}''(x_\mathrm{s}) }{ {\cal G}'(x_\mathrm{s}) }
  {\biggr )},
  \label{diferenca area-luk PCAdS}
 \end{eqnarray}
where $2|{\cal G}''(x_\mathrm{s})| / [{\cal
G}'(x_\mathrm{s})]^2<1$.

Adding (\ref{area BH-luk PCAdS}) and
 (\ref{diferenca area-luk PCAdS}), and using the results of Sec.
\ref{sec:Lukewarm-inst AdS} yields finally the total area of the
nonextreme AdS instanton with $m=q$
\begin{eqnarray}
& & \!\!\!\!\!\!\!\!\!\!{\cal A}_{\rm bh}^{\rm nonext}+\Delta
{\cal A}_{\rm ac}^{\rm nonext} = \nonumber \\
& & -\frac{16\pi m^2}{\omega_+ (\omega_+^2-1)} {\biggl (}\!
 -\frac{\omega_+ -\omega_-}{(\omega_+ +\alpha)(\omega_- +\alpha)}
       +\frac{1}{\omega_+ -\alpha}  \nonumber\\
 & &
 + \frac{1}{ 1+\sqrt{1-\frac{8|\Lambda|m^2}{3}
\frac{3\omega_-^2-1}{\omega_-^2(1-\omega_-^2)^2} }}
 \frac{1-3\omega_-^2}{\omega_-(1-\omega_-^2)} {\biggr )},
 \label{area TOTAL-luk PCAdS}
 \end{eqnarray}
 where $\omega_+$ and $\omega_-$ are defined in (\ref{polos-luk PCAdS}),
 $\alpha$ is given by (\ref{yA-luk PCAdS}),
 and condition (\ref{mq-luk PCAdS}) must be satisfied. The pair
creation rate of nonextreme AdS black holes with $m=q$ is then
\begin{eqnarray}
\Gamma_{\rm nonext}\sim e^{\frac{1}{4}\left (
 {\cal A}_{\rm bh}^{\rm nonext}
 +\Delta {\cal A}_{\rm ac}^{\rm nonext} \right )}.
 \label{rate-luk PCAdS}
 \end{eqnarray}
Fixing $A$ and $\Lambda$ one concludes that the pair creation rate
decreases as the mass of the black holes increases. Moreover,
fixing $m$ and $\Lambda$, in the domain of validity of our
results, $mA\ll 1$ and $A>\sqrt{|\Lambda|/3}$, as $A$ increases
the pair creation rate increases. Hence, the general behavior of
the pair creation rate of nonextreme black holes with $m=q$ in the
AdS case is analogous to the corresponding behavior in the flat
case \cite{HawkRoss-string} (see also Appendix \ref{sec:A1}).

\subsection{\label{sec:Cold-rate AdS}Pair creation rate in the
extreme case ($\bm{y_+=y_-}$)}

In the extreme AdS case, the instanton has a single horizon, the
acceleration horizon at $y=y_A$, in its Euclidean section, since
 $y=y_+$ is an internal infinity.
The pair creation rate of extreme AdS black holes with $y_+=y_-$
is then
\begin{eqnarray}
\Gamma_{\rm ext}\sim e^{\frac{1}{4}
 \Delta {\cal A}_{\rm ac}^{\rm ext}}\,,
 \label{rate-cold PCAdS}
 \end{eqnarray}
where $\Delta {\cal A}_{\rm ac}^{\rm ext}$ is given by
 (\ref{diferenca area-luk PCAdS}), with
 $y_A$ defined in (\ref{zerosy3-cold}), $x_\mathrm{s}$
and $x_\mathrm{n}$ given by (\ref{polos-cold}),
 and condition (\ref{mq-cold}) must be satisfied.
The pair creation rate decreases as the mass of the black holes
increases, and the pair creation rate increases when $A$
increases. The general behavior of the pair creation rate of
extreme black holes as a function of $m$ and $A$ in the AdS case
is also analogous to the behavior of the flat case, discussed in
\cite{HawkRoss-string} (see also Appendix \ref{sec:A1}).

\subsection{\label{sec:submaximal-rate AdS} Pair creation rate in the
nonextreme case with $\bm{m\neq q}$}

The nonextreme instantons, that mediate the pair creation of
nonextreme black holes with $m\neq q$ (including the case $q=0$),
have two horizons in their Euclidean section, namely the
acceleration horizon at $y=y_A$ and the black hole horizon at
$y=y_+$. The pair creation rate of nonextreme AdS black holes with
$m\neq q$ is then
 $\Gamma \sim e^{( {\cal A}_{\rm bh}^{\rm nonext}
 +\Delta {\cal A}_{\rm ac}^{\rm nonext})/4}$, with
${\cal A}_{\rm bh}^{\rm nonext}$ given by
 (\ref{area BH-luk PCAdS}) and $\Delta {\cal A}_{\rm ac}^{\rm nonext}$
given by (\ref{diferenca area-luk PCAdS}), subjected to the
results found in Sec. \ref{sec:submaximal-inst AdS}. The pair
creation rate decreases as the mass of the black holes increases,
and the pair creation rate increases when $A$ increases.

\section{\label{sec:Conc AdS}Summary and discussion}

We have studied in detail the quantum process in which a cosmic
string breaks in an anti-de Sitter (AdS) background and a pair of
black holes is created at the ends of the string. The energy to
materialize and accelerate the black holes comes from the strings
tension. The analysis of this process in a flat background
($\Lambda=0$) has been carried in \cite{HawkRoss-string}, while in
a de Sitter background ($\Lambda>0$) it has been done in
\cite{OscLem-PCdS}. In an AdS background this is the only study
done in the process of production of a pair of correlated black
holes with spherical topology. We remark that in principle our
explicit values for the pair creation rates also apply to the
process of pair creation in an external electromagnetic field,
with the acceleration being provided in this case by the Lorentz
force instead of being furnished by the string tension. Indeed,
there is no AdS Ernst solution, and thus we cannot discuss
analytically the process. However, physically we could in
principle consider an external electromagnetic field that supplies
the same energy and acceleration as our strings and, from the
results of the $\Lambda=0$ case (where the pair creation rates in
the string and electromagnetic cases agree), we expect that the
pair creation rates found in this paper do not depend on whether
the energy is being provided by an external electromagnetic field
or by a string.

It is well known that the AdS background is attractive, i.e., an
analysis of the geodesic equations indicates that particles in
this background are subjected to a potential well that attracts
them (see also Appendix \ref{sec:A2}). Therefore, if we have a
virtual pair of black holes and we want to turn them real, we will
have to furnish a sufficient force that overcomes this
cosmological background attraction. We then expect that pair
creation is possible only if the strings tension and the
associated acceleration $A$ is higher than a critical value. We
have confirmed that this is indeed the case: in the AdS
background, black holes can be pair produced only with an
acceleration higher than $\sqrt{|\Lambda|/3}$.

We have constructed the saddle point solutions that mediate the
pair creation process through the analytic continuation of the AdS
C-metric, and we have explicitly computed the pair creation rate
of the process. The AdS pair creation rate reduces to the
corresponding one of the flat case \cite{HawkRoss-string} when we
set $\Lambda=0$. We have concluded that, for a pair of black holes
that is subjected to a fixed $\Lambda$ and $A$ backgrounds, the
pair creation probability decreases when the mass or charge of the
black holes increases. Moreover, when we fix the mass and the
charge of the black holes, the probability they have to be pair
created increases when the acceleration provided by the string
increases. These results are physically expected, as an heuristic
derivation done in Appendix \ref{sec:A2} confirms. This process
has also a clear analogy with a thermodynamical system, with the
mass density of the string (that is proportional to $A$) being the
analogue of the temperature $T$. Indeed, from the Boltzmann
factor, $e^{-E_0/(k_{\rm B} T)}$ (where $k_{\rm B}$ is the
Boltzmann constant), one knows that a higher background
temperature makes the nucleation of a particle with energy $E_0$
more probable. Equivalently, a higher acceleration provided by the
string makes the creation of the black hole pair more probable.

\begin{figure} [h]
\includegraphics*[height=5in]{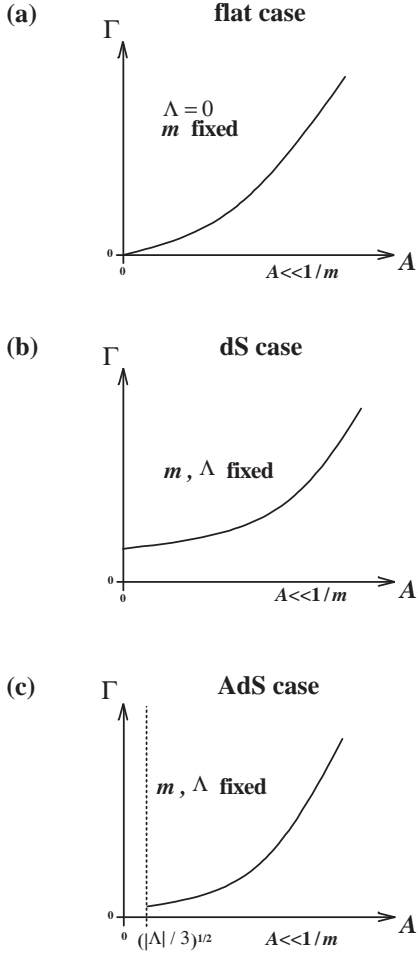}
\caption{\label{PC rates_AdS}
 General behavior of the black hole pair creation rate $\Gamma$ as a function
of the acceleration $A$ provided by the strings, when a cosmic
string breaks: (a)  in a flat background ($\Lambda=0$)
\cite{HawkRoss-string}, (b)  in a dS background ($\Lambda>0$)
\cite{OscLem-PCdS}, and (c)  in an AdS background ($\Lambda<0$).
 }
\end{figure}

For the benefit of comparison, in Fig. \ref{PC rates_AdS} we
schematically represent the general behavior of the black hole
pair creation rate $\Gamma$ as a function of the acceleration $A$
provided by the strings, when a cosmic string breaks in the three
cosmological constant backgrounds. In a flat background [see Fig.
\ref{PC rates_AdS}.(a)], the pair creation rate is zero when $A=0$
\cite{HawkRoss-string}. In this case, the flat C-metric reduces to
a single black hole, and since we are studying the probability of
pair creation, the corresponding rate is naturally zero. This does
not mean that a single black hole cannot be materialized from the
quantum vacuum, it only means that this latter process is not
described by the C-metric. The creation probability of a single
black hole in a hot bath has been considered in
\cite{GrossPerryYaffe}. In a dS background [see Fig. \ref{PC
rates_AdS}.(b)], the pair creation rate is not zero when $A=0$
\cite{OscLem-PCdS}. This means that even in the absence of the
string, the positive cosmological constant is enough to provide
the energy to materialize the black hole pair \cite{MannRoss}. If
in addition one has an extra energy provided by the string, the
process becomes more favorable \cite{OscLem-PCdS}. In the AdS case
[see Fig. \ref{PC rates_AdS}.(c)], the negative cosmological
constant makes a negative contribution to the process, and black
hole pair creation is possible only when the acceleration provided
by the strings overcomes the AdS background attraction. The branch
$0<A\leq \sqrt{|\Lambda|/3}$ represents the creation probability
of a single black hole when the acceleration provided by the
broken string is not enough to overcome the AdS attraction, and
was not studied in this paper.

We have also verified that (as occurs with pair creation in other
backgrounds) the pair production of nonextreme black holes is
enhanced relative to the pair creation of extreme black holes by a
factor of $e^{S_{\rm bh}}$, where $S_{\rm bh}={\cal A}_{\rm bh}/4$
is the gravitational entropy of the black hole.


\begin{acknowledgments}

It is a pleasure to acknowledge conversations with Jos\'e P. S.
Lemos, Vitor Cardoso,
 and Alfredo B. Henriques. This work was
partially funded by Funda\c c\~ao para a Ci\^encia e Tecnologia
(FCT) through project CERN/FIS/43797/2001 and PESO/PRO/2000/4014.
I also acknowledge finantial support from the FCT through PRAXIS
XXI programme.

\end{acknowledgments}

\appendix
\section{\label{sec:A1}Pair creation rate in the flat case}
In a flat background ($\Lambda=0$), the analysis of the process of
pair creation of black holes when a cosmic string breaks has been
analyzed by Hawking and Ross \cite{HawkRoss-string}. In this case,
there is a direct $\Lambda=0$ counterpart of the nonextreme AdS
instanton and of the extreme AdS instanton discussed in our AdS
case. These instantons describe pair creation of nonextreme (with
$m=q$) and extreme black holes with $y_+=y_-$, respectively
\cite{HawkRoss-string}. The total area of these $\Lambda=0$
instantons can be found in \cite{HawkRoss-string}, and can be
obtained by taking the direct $\Lambda=0$ limit of (\ref{area
BH-luk PCAdS}) and (\ref{diferenca area-luk PCAdS}), together with
the replacement $y_A \mapsto -x_{\mathrm s}$. This procedure
yields that (\ref{area BH-luk PCAdS}) also holds in the
$\Lambda=0$ case, while  $\Delta {\cal A}_{\rm ac}$ becomes
\begin{eqnarray}
\!\!\!\!\!\Delta {\cal A}_{\rm ac} =
   -\frac{4\pi} {A^2 |{\cal G}'(x_\mathrm{n})|}\,
 {\biggl (} \frac{1}{ x_\mathrm{n}-x_{\mathrm
s}}
 + \frac{1}{2}
 \frac{ {\cal G}''(x_\mathrm{s}) }{ {\cal G}'(x_\mathrm{s}) }
  {\biggr )} \,.
  \label{diferenca area-flat}
 \end{eqnarray}
In \cite{HawkRoss-string} the explicit numerical value of ${\cal
A}_{\rm bh}$ and $\Delta {\cal A}_{\rm ac}$ has not been computed.
We will do it here.

For the $\Lambda=0$ nonextreme case, the discussion of Sec.
\ref{sec:Lukewarm-inst AdS} applies generically, as long as we set
$\Lambda=0$ in the corresponding equations. With this data we find
the explicit value of the total area of the nonextreme flat
instanton
\begin{eqnarray}
 \!\!\!\!\!\!\!\!\!\!\!\!& &{\cal A}_{\rm bh}+\Delta
{\cal A}_{\rm ac} =
-\frac{\pi}{A^2}\frac{-1+4mA+\sqrt{1-(4mA)^2}}{\sqrt{1-(4mA)^2}}.\nonumber \\
& &
 \label{area TOTAL-flat}
 \end{eqnarray}
In the particle limit, $mA\ll 1$, the above relation reduces to
$-\pi m/(4A)$ and the mass density of the string is given by
$\mu\sim mA$. The pair creation rate is then $\Gamma\sim e^{-\pi
m^2/\mu}$. Thus, as occurs with the AdS case, the pair creation
rate decreases when $m$ increases and the rate increases when $A$
or $\mu$ increase [see Fig. \ref{PC rates_AdS}.(a)].

We remark that for $mA\sim 1$, and as occurs in the corresponding
AdS case, the pair creation rate associated to (\ref{area
TOTAL-flat}) starts decreasing when $A$ increases. This is a
physically unexpected result since a higher acceleration provided
by the string background should favor the nucleation of a fixed
black hole mass. The sector $mA\sim 1$ must then be discarded and
the reason is perfectly identified: the domain of validity of the
rates is $mA\ll 1$, for which the radius of the black hole, $r_+
\sim m$, is much smaller than the typical distance between the
black holes at the creation moment, $\ell \sim 1/A$ (this value
follows from the Rindler motion $x^2-t^2=1/A^2$ that describes the
uniformly accelerated motion of the black holes). So, for $mA\sim
1$ one has $r_+\sim \ell$ and the black holes start interacting
with each other.

In what concerns the pair creation rate of extreme and nonextreme
with $\bm{m=q}$ $\Lambda=0$ black holes, an explicit computation
shows that its general behavior with $A$ and $m$ is also similar
to the one of the AdS case, i.e., the rate decreases when $m$ or
$q$ increase, and the rate increases when $A$ increases.

\section{\label{sec:A2}Heuristic derivation of the pair creation
rates}
In order to clarify the physical interpretation of the results, in
this Appendix we heuristically derive some results discussed in
the main body of the paper. In particular, we find heuristically
the pair creation rates.

In an AdS background, pair creation of black holes is possible
only when the acceleration provided by the strings satisfies
$A>\sqrt{|\Lambda|/3}$. To understand this result one can argue as
follows. In general, the time-time component of the gravitational
field is given by $g_{00}=1-2\Phi$, where $\Phi$ is the Newtonian
potential. In the AdS spacetime, one has $\Phi=-|\Lambda| r^2/6$
and its derivative yields the force per unit mass or acceleration
of the AdS spacetime, $A_{\rm AdS}=-|\Lambda|r/3 \sim
-\sqrt{|\Lambda|/3}$, where we have replaced $r$ by the
characteristic AdS radius $\left (|\Lambda|/3 \right )^{-1/2}$.
The minus sign indicates that the AdS background is attractive and
thus, if one wants to have a pair of accelerated black holes
driving away from each other, the cosmic string will have to
provide a sufficient acceleration $A$ that overcomes the AdS
background attraction, i.e., $A>|A_{\rm AdS}|$.

An estimate for the black hole pair creation probability can be
given by the Boltzmann factor, $\Gamma \sim e^{-E_0/W_{\rm ext}}$,
where $E_0$ is the energy of the system that nucleates and $W_{\rm
ext}=F \ell$ is the work done by the external force $F$, that
provides the energy for the nucleation, through the typical
distance $\ell$ separating the created pair. First we ask what is
the probability that a black hole pair is created in a $\Lambda=0$
background when a string breaks. This process has been discussed
in \cite{HawkRoss-string} (see also Appendix \ref{sec:A1}) where
it was found that the pair creation rate is $\Gamma \sim
e^{-m/A}$. In this case, $E_0 \sim 2m$, where $m$ is the rest
energy of the black hole, and $W_{\rm ext}\sim A$ is the work
provided by the strings. To derive $W_{\rm ext}\sim A$ one can
argue as follows. The  acceleration provided by the string is $A$,
the characteristic distance that separates the pair at the
creation moment is $1/A$ (see Appendix \ref{sec:A1}), and the
characteristic mass of the system is A by the Compton relation.
Thus, the characteristic work is $W_{\rm ext}={\rm mass}\times{\rm
acceleration}\times{\rm distance}\sim A A A^{-1}$. So, from the
Boltzmann factor we indeed expect that the creation rate of a
black hole pair when a string breaks in a $\Lambda=0$ background
is given by $\Gamma \sim e^{-m/A}$.

Now we ask what is the probability that a string breaks in an AdS
background and a pair of black holes is produced. As we saw just
above, the presence of the AdS background leads in practice to a
problem in which we have a net acceleration that satisfies
$A'\equiv \sqrt{A^2-|\Lambda|/3}$, this is, $\Lambda$ makes a
negative contribution to the process. Heuristically, we may then
apply the same arguments that have been used in the last
paragraph, with the replacement $A\rightarrow A'$. At the end, the
Boltzmann factor tells us that the creation rate for the process
is $\Gamma \sim e^{-m/\sqrt{A^2-|\Lambda|/3}}$. So, given  $m$ and
$\Lambda$, when the acceleration provided by the string grows the
pair creation rate increases, as the explicit calculations done in
the main body of the paper show.


\end{document}